\documentclass[journal=ancac3,manuscript=article]{achemso} 

\usepackage{amsmath}
\usepackage{graphicx}
\usepackage{siunitx}

\author{Jiří Kabát}
\affiliation[CEITEC]
{Brno University of Technology, Central European Institute of Technology, Purky\v{n}ova 123, 612 00, Brno, Czech Republic}
\alsoaffiliation[UFI]
{Brno University of Technology, Faculty of Mechanical Engineering, Institute of Physical Engineering, Technická 2, 616 69, Brno, Czech Republic}

\author{Rostislav Řepa}
\affiliation[CEITEC]
{Brno University of Technology, Central European Institute of Technology, Purky\v{n}ova 123, 612 00, Brno, Czech Republic}
\alsoaffiliation[UFI]
{Brno University of Technology, Faculty of Mechanical Engineering, Institute of Physical Engineering, Technická 2, 616 69, Brno, Czech Republic}

\author{Jordan A. Hachtel}
\affiliation[ORNL]
{Center for Nanophase Materials Sciences, Oak Ridge National Laboratory, Oak Ridge, TN 37830, USA}

\author{Peter Kepič}
\affiliation[CEITEC]
{Brno University of Technology, Central European Institute of Technology, Purky\v{n}ova 123, 612 00, Brno, Czech Republic}

\author{Vlastimil Křápek}
\affiliation[CEITEC]
{Brno University of Technology, Central European Institute of Technology, Purky\v{n}ova 123, 612 00, Brno, Czech Republic}
\alsoaffiliation[UFI]
{Brno University of Technology, Faculty of Mechanical Engineering, Institute of Physical Engineering, Technická 2, 616 69, Brno, Czech Republic}

\author{Andrea Konečná}
\affiliation[CEITEC]
{Brno University of Technology, Central European Institute of Technology, Purky\v{n}ova 123, 612 00, Brno, Czech Republic}
\alsoaffiliation[UFI]
{Brno University of Technology, Faculty of Mechanical Engineering, Institute of Physical Engineering, Technická 2, 616 69, Brno, Czech Republic}

\author{Tomáš Šikola}
\affiliation[CEITEC]
{Brno University of Technology, Central European Institute of Technology, Purky\v{n}ova 123, 612 00, Brno, Czech Republic}
\alsoaffiliation[UFI]
{Brno University of Technology, Faculty of Mechanical Engineering, Institute of Physical Engineering, Technická 2, 616 69, Brno, Czech Republic}

\author{Michal Horák}
\affiliation[CEITEC]
{Brno University of Technology, Central European Institute of Technology, Purky\v{n}ova 123, 612 00, Brno, Czech Republic}
\email{michal.horak2@ceitec.vutbr.cz}

\title{Tuning of Localized Surface Plasmons\\in Vanadium Dioxide Nanoparticles\\via Size and Insulator--Metal Transition}

\begin{document}

\begin{abstract}

Vanadium dioxide has been identified as a promising phase-changing material for use in tunable plasmonic devices. In this study, we present a comprehensive modal analysis of single-phase and multi-phase vanadium dioxide nanoparticles. In-situ high-resolution electron energy loss spectroscopy was utilized to experimentally resolve the dipole plasmon peak, higher-order and breathing plasmonic modes, and bulk losses as a function of nanoparticle size. Furthermore, the focus is directed toward capturing the dynamic nanoscale optical response throughout the metal-insulator transition in a vanadium dioxide nanoparticle. This system possesses the ability to be gradually switched on and off in terms of the emergence of near-infrared plasmonic absorption. The switching is accompanied by a gradual spectral shift of the absorption peak, amounting to 0.18 eV for a 120 nm nanoparticle. It is envisioned that this phenomenon can be generalized to larger nanostructures with a higher aspect ratio, thereby introducing a wider tunability of the system, which is essential for functional nanodevices based on vanadium dioxide.

\end{abstract}

% =========================================================
% INTRODUCTION
% =========================================================

\noindent\hrulefill

\section{Introduction}

Metallic nanostructures have been utilized to enhance the efficiency of photovoltaic devices~\cite{Ai2022}, catalytic reactions~\cite{Liu2025}, single-photon emitters~\cite{Zhu2016}, and even the detection of individual molecules~\cite{Altug2022}. This improvement is made possible by localized surface plasmon resonances in metallic nanostructures, which significantly enhance electric field, absorption, or temperature at the nanoscale~\cite{Novotny2006_NanoOptics, Maier2007}. The most widely used metals in these nanostructures are gold, silver, or aluminum, which exhibit the best metallic properties (the lowest negative real and simultaneously the lowest positive imaginary parts of the dielectric function)~\cite{West2010}. However, the function of nanostructures made of these metals remains fixed and cannot be tuned after fabrication. Naturally, making the function of plasmonic devices tunable with external stimuli, especially on an ultrashort timescale, would be attractive for many applications~\cite{Lian2022}. One of the most effective approaches for tunable plasmonic devices is the fabrication of nanostructures from phase-change materials~\cite{Youngblood2023}.

Among these phase-change materials, vanadium dioxide (VO$_2$) stands out due to the insulator--metal phase transition, which occurs at a convenient temperature of around \SI{67}{\celsius}~\cite{Morin1959, Cueff2020}. As the transition arises not only from the Peierls-based change from monoclinic to rutile crystallographic structure~\cite{Eyert2002}, but also from the Mott-Hubbard-based transition of correlated electrons~\cite{Wentzcovitch1994}, it can be triggered electrically~\cite{Markov2015, Kim2010} or optically on a \SI{e2}{fs} time scale~\cite{Lopez20041, Hallman2021}. Both the speed and the energy of the transition made VO$_2$ attractive for integrated circuits~\cite{Parra2021, Jung2021, Jung2022} or free-space devices~\cite{Kepic2021, King2024}. Even more applications have arisen since the transition hysteresis behavior and its timescale have been studied within individual crystal grains using X-ray~\cite{Johnson2022, Jager2017} or electron microscopy techniques~\cite{Otto2019, Fu2020, Sood2021, Kim2023}. Although providing valuable insights into the phase transition of VO$_2$, these studies were primarily based on X-ray absorption or electron diffraction, which reflect structural rather than electrical or optical changes that are more relevant to applications.

In our recent studies of individual VO$_2$ nanoparticles (NPs) using transmission electron microscopy techniques~\cite{Kepic2025}, the transition and hysteresis were also probed by tracking plasmonic resonances that appear when VO$_2$ undergoes a transition from an insulator to a metal. Despite being able to link those results directly to optical properties, these plasmonic resonances and their modes have been discussed sparsely. Besides our study, localized surface plasmon resonances have been investigated in detail only in VO$_2$ nanostructure arrays using optical microscopy techniques~\cite{Lopez2002, Lopez20042, Lei2010, Appavoo2012, Butakov2017, Kepic2021}, which did not provide comprehensive information about individual plasmonic modes and their tunability.

Here, we present a spatial modal analysis of excitations within individual VO$_2$ nanoparticles in both the low-temperature insulating and high-temperature metallic phases using in-situ scanning transmission electron microscopy (STEM) with electron energy-loss spectroscopy (EELS). STEM EELS has been shown to be a powerful technique for characterizing and studying optical modes within individual plasmonic nanostructures~\cite{Abajo2010, Colliex2016, Horak2023}. That is why we utilize high-resolution EELS in order to experimentally resolve the dipole plasmon peak, in addition to potentially higher-order and breathing plasmonic modes and bulk losses. This is achieved through straightforward post-processing using Lorentzian profile fitting with only two amplitude-free parameters, thus providing a full modal analysis of both insulating and metallic VO$_2$ NPs. As a result, we introduce an experimental observation of the energy of the dipole plasmon mode and its other parameters (amplitude and quality factor) as a function of the size of the metallic VO$_2$ NPs. Furthermore, we focus on transitional states within the insulator-to-metal transition. The dynamics of the phase transition facilitated the specific particle's ability to establish a correlation between the local temperature and the relative switched volume. This correlation was achieved by employing the dipole plasmon mode in the metallic region as the linking variable, with the peak position serving as the indicator.

% =========================================================
% Results and Discussion
% =========================================================

\section{Results and Discussion}

\begin{figure}[t!]
\includegraphics[width=1\textwidth]{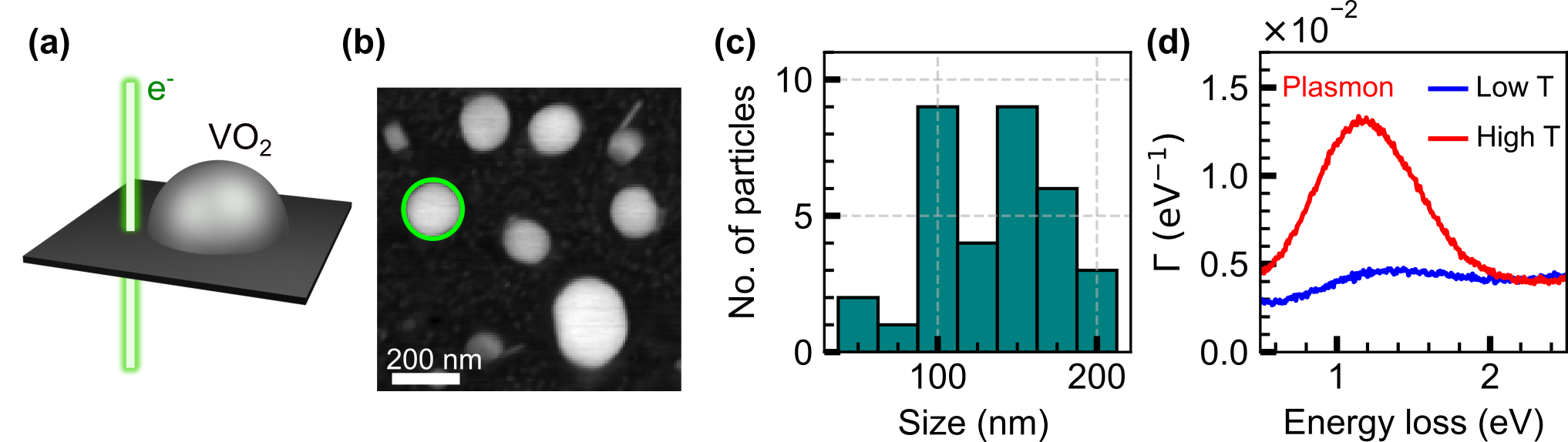}
\caption{Vanadium dioxide nanoparticles. (a) Schematic of the hemispherical VO$_2$ NP on a SiN$_x$ membrane with the incident electron beam. (b) HAADF STEM image of one region of an ensemble of VO$_2$ NPs. (c) Size distribution of the whole ensemble of NPs. (d) Measured EEL spectra of \SI{170}{\nano\metre} NP (highlighted in (b) with a green circle) at low (\SI{25}{\celsius}) and high (\SI{180}{\celsius}) temperature. The particle switches from an insulating (blue) phase to a metallic (red) phase. In the metallic phase, a pronounced plasmon peak emerges. The EEL spectra are background-subtracted and integrated over the whole particle. 
\label{fig1}}
\end{figure}

Vanadium dioxide nanoparticles (VO$_2$ NPs) of hemispherical shape were prepared on a thin silicon nitride membrane by a two-step process consisting of thin film dewetting, as introduced in Ref. \cite{Kepic2025} (for more details, see Methods). Figure~\ref{fig1}a shows a high-angle annular dark field (HAADF) STEM image of VO$_2$ NPs. Their size was evaluated from several regions of the sample and ranges from 50 to \SI{220}{\nano\metre}, as shown in Fig.~\ref{fig1}b. In the following, we focus on the \SI{170}{\nano\metre} VO$_2$ NP to show the typical characteristics of low-loss EEL spectra (Fig.~\ref{fig1}c). The EEL spectra are integrated over the whole nanoparticle, the background is subtracted, and the energy-dependent loss probability $\Gamma(E)$ is plotted. At \SI{25}{\celsius}, the VO$_2$ NP is in the insulating phase, and there is no plasmon peak in the low-loss EELS (blue line). When heated to \SI{180}{\celsius}, it undergoes a phase change to the metallic phase, and a plasmon peak appears around \SI{1.2}{\electronvolt} (red line), which is the well-known characteristic of the VO$_2$ NPs \cite{Kepic2025}. However, the plasmonic response has not yet been thoroughly analyzed. Thus, we perform a complete modal analysis of insulating, fully metallic and partially metallic VO$_2$ NPs. STEM EELS enables us to analyze the spatial dependence of near-field excitations, such as localized surface plasmons and even a bulk plasmon, which is inaccessible to light microscopy. Subsequently, we focus on the \SI{170}{\nano\metre} VO$_2$ NP in the low-temperature insulating phase and the high-temperature metallic phase to explain all the characteristics of the EEL spectra.

\subsection{Single-phase VO$_2$ nanoparticles}

\begin{figure}[t!]
\includegraphics[width=1\textwidth]{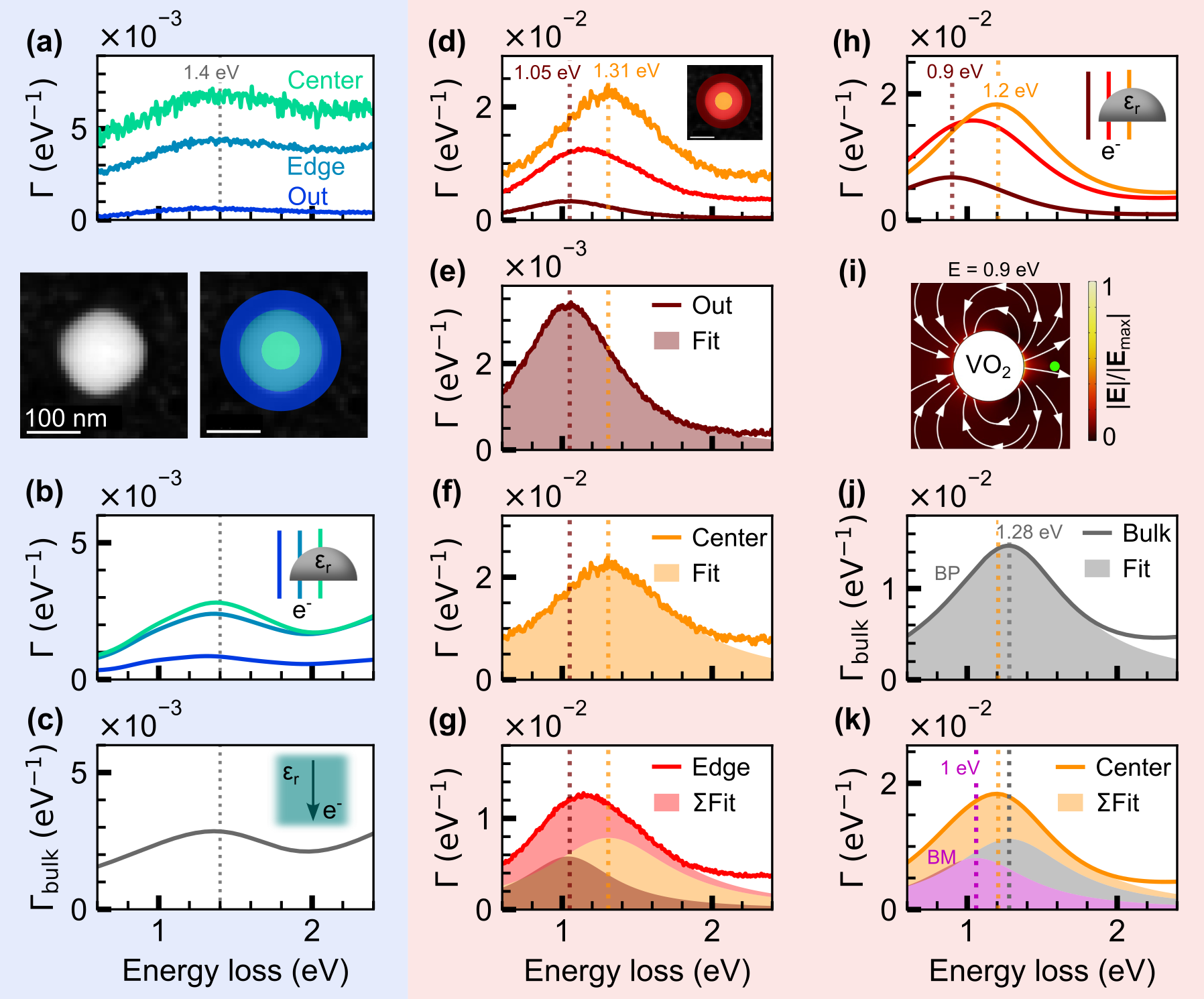}
\caption{Spatial modal analysis of \SI{170}{\nano\metre} VO$_2$ NP. (a-c) Insulating VO$_2$ NP. (a) Measured EEL spectra for the insulating VO$_2$ NP (at \SI{25}{\celsius}) at the three regions (center, edge, out; highlighted next to the HAADF STEM micrograph of the VO$_2$ NP). (b) Numerically simulated EEL spectra for the corresponding three electron beam positions. (c) Calculated loss function of insulating bulk VO$_2$. (d-k) Metallic VO$_2$ NP. (d) Measured EEL spectra for the metallic VO$_2$ NP (at \SI{180}{\celsius}) from the same three regions as in (a). (e,f) Determination of the dipole mode (e) and bulk plasmon (f) energy. (g) Processing of the EEL spectrum integrated over the edge region by two Lorentzians corresponding to the dipole and bulk plasmon. (h) Numerically simulated EEL spectra. (i) Normalized magnitude of the induced electric field with the field lines around the NP at the energy \SI{0.9}{\electronvolt} resembling the in-plane dipole field. Electron beam passing \SI{70}{\nano\metre} outside the particle is marked with the green dot. (j) Calculated loss function of metallic bulk VO$_2$. (k) Processing of the calculated EEL spectrum in the center of the NP by two Lorentzians, one with fixed peak energy and FWHM, corresponding to the bulk loss function, and one with free parameters, corresponding to the breathing mode (BM).
\label{fig2}}
\end{figure}

First, we investigate the EEL spectra of insulating VO$_2$ NP (Fig.~\ref{fig2}a-c). Figure~\ref{fig2}a shows the HAADF STEM micrograph of a \SI{170}{\nano\metre} NP and the HAADF STEM image with three highlighted regions over which we integrate the EEL signal to obtain the spectra. Note that the processing of EEL spectra is shown in detail in the Supplementary Information (Figure~\ref{SI_fig1}). The outermost region (out, blue) is placed outside the NP to exclude bulk contributions of the insulating VO$_2$ to the EEL spectra. The second region takes the signal from the inside edge of the NP (edge, teal), and the innermost region takes the signal from the center (center, green). The background-subtracted experimentally measured EEL spectra are plotted in Fig.~\ref{fig2}a with the color coding matching the regions from which the spectra were taken. We observe a decreasing loss probability when moving further away from the center, with a major drop when the spectra were taken outside. The peak originating around \SI{1.4}{\electronvolt} is attributed to material losses, particularly the interband transitions in the insulating VO$_2$~\cite{Eyert2002}.

In the following, we compare the experimental spectra with the simulations for a hemispherical VO$_2$ NP with a diameter of \SI{170}{\nano\metre} (for simulation details, see Methods). To match the experiment, the following positions of the electron beam were used: outside the NP \SI{10}{\nano\metre} from its edge (out, blue), inside the NP \SI{10}{\nano\metre} from its edge (edge, teal), and in the center of the NP (center, green). The calculated spectra in Fig.~\ref{fig2}b reveal similar features as the experimental ones. In addition, we verify that the major contribution to the EEL spectra originates from the bulk material absorption by calculating the bulk loss function (see Methods, Eq.~\eqref{eq_Bulk_loss}) for an electron passing through \SI{85}{\nano\metre} of a bulk insulating VO$_2$, which corresponds to the height of the hemisphere. The calculated spectrum is shown in Fig.~\ref{fig2}c with a peak at \SI{1.4}{\electronvolt}.

Second, we focus on the EEL spectra of metallic VO$_2$ NP (Fig.~\ref{fig2}d-k). The EEL spectra in Fig.~\ref{fig2}d are extracted from the same three regions, as for the insulating phase: the outermost region (out, maroon), inside the nanoparticle near its edge (edge, red), and the innermost region (center, orange). Similarly to the case of insulating VO$_2$, we observe a decrease in loss probability when moving further away from the center, with a major drop when the spectra were taken outside. Furthermore, we observe a spectral shift of the peak. To quantify the change in peak height, spectral position, and width, we fit the spectra with the Lorentzian line profile 
\begin{align}
\Gamma(E) = \frac{A\gamma^2}{(E-E_\mathrm{peak})^2+\gamma^2},\label{Lorentz}
\end{align}
where $A$ is the peak height, $E_\mathrm{peak}$ is the spectral position of the peak, and $\gamma$ is the damping parameter, related to the full width at half maximum (FWHM), which is equal to $2\gamma$.

The results of the fit are summarized in Fig.~\ref{fig2}e-g. First, we focus on the outer region (Fig.~\ref{fig2}e) with a central energy E$_\mathrm{peak}$ of \SI{1.05}{\electronvolt}. The fact that the peak emerges even outside the nanoparticle indicates that there are losses due to the near-field excitation of localized surface plasmons (LSP) within the VO$_2$ NP. Concerning the peak excited for a beam position in the center of the nanoparticle (Fig.~\ref{fig2}f), the energy reads \SI{1.31}{\electronvolt}. Due to the position of the beam, it is assumed that the peak consists of bulk (material) losses or contributions of plasmonic modes other than the in-plane dipole mode (e.g., breathing mode). Lastly, we fit the edge region (Fig.~\ref{fig2}g), where both the LSP and the bulk contributions are expected. We therefore fit the spectrum with the sum of two Lorentzian curves with fixed energies \SI{1.05}{\electronvolt} and \SI{1.31}{\electronvolt}. The shaded areas under the curves highlight both the sum (red) and individual contributions of LSP (maroon) and bulk plasmon (orange). In traditional plasmonic materials like gold or silver, one expects the bulk, surface, and LSP contributions to be well separated in energy. However, due to their large width caused by extensive damping in VO$_2$, the peaks overlap. We note that to exclude the spectral overlap of the LSP contribution and the bulk plasmon peak, larger structures are needed.

We again compare the experimental spectra with the simulations with the following beam positions: center of the NP (center, orange), inside the NP \SI{10}{\nano\metre} from its edge (edge, red), and outside the NP \SI{10}{\nano\metre} from its edge (out, maroon). The calculated spectra in Fig.~\ref{fig2}h show a behavior similar to the experimental, with the energy of the LSP \SI{0.9}{\electronvolt} and the peak excited for the beam position in the center at \SI{1.2}{\electronvolt}. The discrepancy between the experiment and the simulation can be most probably attributed to a difference in the dielectric function $\varepsilon_\mathrm{r}$ used in the model, which was measured for a thin polycrystalline VO$_2$ film~\cite{Kepic2025} and the realistic $\varepsilon_\mathrm{r}$ of a single crystal VO$_2$ NP. In addition, we plot the magnitude of the induced electric field (electron beam positioned \SI{70}{\nano\metre} from the outer edge) with the field lines in Fig.~\ref{fig2}i, at the loss energy \SI{0.9}{\electronvolt}. We can see that it resembles a field of an in-plane dipole, slightly distorted by the electron beam. Consequently, the peak at \SI{0.9}{\electronvolt} corresponds to the dipole LSP mode. 

Next, we elucidate the origin of the peak at \SI{1.2}{\electronvolt}. We expect that while the beam passes through the center of the NP, the bulk plasmon is excited. We verify it by calculating the bulk loss function of the metallic VO$_2$. The calculated spectrum is shown in Fig.~\ref{fig2}j where a peak is present at the energy of \SI{1.28}{\electronvolt} corresponding to the bulk plasmon. The energy is slightly higher than in the simulations for the center position (\SI{1.2}{\electronvolt}). We attribute the shift to a possible excitation of another LSP mode, primarily the breathing mode, owing to the center position of the electron beam. We fit the spectrum for the center position (orange line in Fig.~\ref{fig2}h) with two Lorentzian curves. The result of the fit is presented in Fig.~\ref{fig2}k. One of them (gray) corresponds to the bulk plasmon with two fixed parameters $E_\mathrm{peak}=$~\SI{1.28}{\electronvolt}, $\gamma=$~\SI{0.48}{\electronvolt}, obtained from the Lorentzian fit of the analytic calculation in Fig.~\ref{fig2}j. The second Lorentzian function (magenta), with all fit parameters set to free, yielded an additional peak at the energy \SI{1}{\electronvolt}. The latter corresponds to the breathing mode. However, because of the prevalence of the bulk plasmon contribution, we will refer to the loss peak from the center as the bulk plasmon.

\begin{figure}[t!]
\includegraphics[width=1\textwidth]{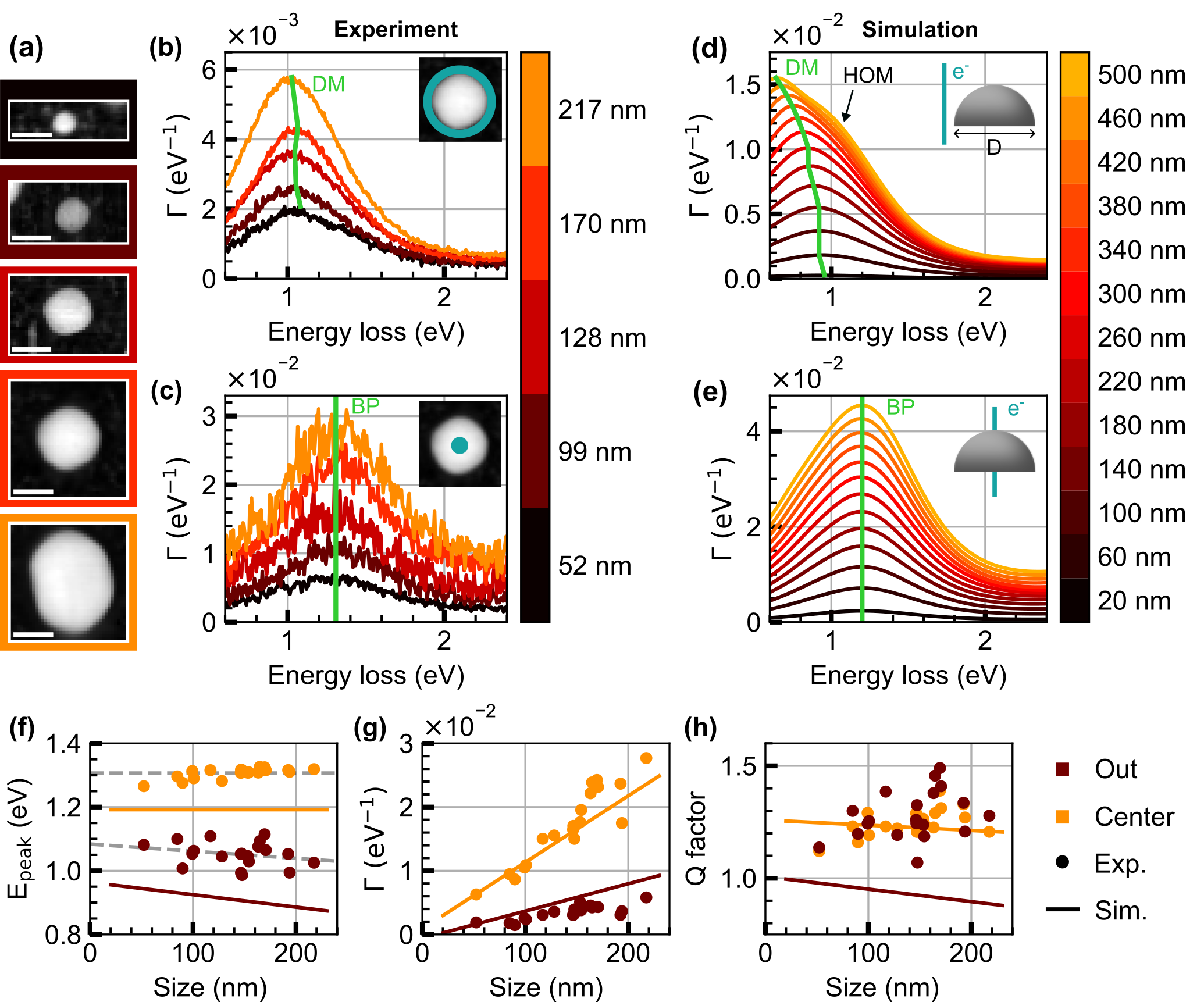}
\caption{Size dependence of EEL spectra for metallic VO$_2$ NPs. (a) HAADF STEM microcrographs of five representatives. The scalebars correspond to \SI{100}{\nano\metre}. (b) Measured EEL spectra taken from the region outside the NPs, where the LSP (dipole mode -- DM) is excited. (c) Measured EEL spectra taken from the center of the NPs, where the bulk plasmon (BP) is excited. (d) Simulated EEL spectra for the beam position \SI{10}{\nano\metre} from the outer edge of the NPs. We observe the emergence of higher-order modes (HOM) for the largest NPs. (e) Simulated EEL spectra for the electron beam positioned in the center. The sizes of the NPs are at the individual colorbar legends. (f) Peak energy, (g) peak loss probability, and (h) Q factor of the dipole (maroon) and bulk (orange) plasmon mode. The experimental energies in (f) are fitted with silver dashed lines as a guide for the eyes. 
\label{fig3}}
\end{figure}

So far, we have explained the observed features in the EEL spectra for the single \SI{170}{\nano\metre} NP. In the following, we focus on the plasmonic contributions to the EEL spectra for different sizes of metallic VO$_2$ NPs from the measured ensemble. Figure~\ref{fig3}a shows HAADF STEM images of five selected NPs of sizes ranging from \SI{50}{\nano\metre} to \SI{220}{\nano\metre}. To observe the change in the plasmonic response with the size of the NP, we chose two regions from which we acquire the EEL signal. Figure~\ref{fig3}b shows the EEL spectra taken from the region outside the edge of the particles, where the LSP excitation occurs without bulk losses. We observe an increase in the signal with particle size and a slight shift of the peak towards lower energies, indicated by the green line. Looking at the central position in Fig.~\ref{fig3}c, we observe a signal increase with the size of the NP. The peak energy is constant at \SI{1.31}{\electronvolt}, indicated by the green line, as expected for the bulk material excitation. For completeness, energy filtered loss probability maps of an ensemble of metallic VO$_2$ NPs at the energy of \SI{0.85}{\electronvolt} and \SI{1.30}{\electronvolt} are shown in Fig.~\ref{SI_fig2}. At the lower energy, the highest loss probability is detected close to the edges of the NPs, indicating the dominant contribution of localized surface plasmons (dipole mode), whereas at the higher energy, the hotspots of the loss probability in the center of the NPs confirm the dominant excitation of bulk plasmon.

A subsequent comparison is made between the measured spectra and numerical simulations. We chose a wide range of nanoparticle sizes ranging from \SI{20}{\nano\metre} to \SI{500}{\nano\metre}. The calculated EEL spectra for the position of the electron beam outside, \SI{10}{\nano\metre} from the edge of the NP, are shown in Fig.~\ref{fig3}d. We observe a pronounced red-shift of the dipole mode (DM) LSP peak with increasing NP size, spanning \SI{0.32}{\electronvolt}, indicated by the green line. In addition, we also observe the possible excitation of higher-order modes (HOM), as shoulders appearing in the loss probability at energies close to \SI{1}{\electronvolt} for the largest NPs. When the electron beam is positioned in the center of the NP, see Fig.~\ref{fig3}e, we observe no peak shift, with the central energy being \SI{1.2}{\electronvolt} (green line), close to the bulk-plasmon energy.

We note that we plot the equivalent comparison between the experimental and simulated data for different NP sizes in the insulating phase in Fig.~\ref{SI_fig3}. The spectra exhibit behavior similar to that of the metallic ones. For the center position, we observe an increase in the signal. In the case of beam passing outside the NPs,  we observe an increase in the signal, and for the largest NPs, also a spectral shift of the peaks. This shift may indicate additional excitations of localized resonances in the dielectrics.

To quantify peak shifts, intensity changes, and quality factors of the resonances for the whole ensemble of NPs, we fit the spectra presented in Fig.~\ref{fig3}b-e by Lorentzian profiles (as introduced in Fig.~\ref{fig2}) for two regions (one in the center, one outside the edge). First, we examine the energy shifts (Fig.~\ref{fig3}f). The energy of the peaks for the center position is constant for different sizes of NPs for both the experiment with an average energy of \SI{1.31}{\electronvolt} (orange dots with a silver dashed line as a guide for the eye) and the simulation with an average energy of \SI{1.2}{\electronvolt} (orange line). For the position outside the edge, we observe a red shift of the peak with increasing the size of the NPs for both the experiment (maroon dots fitted with a silver dashed line as a guide for the eye) and the simulations (maroon line). We note that both bulk plasmon and LSP energies are lower for simulations compared to the experiment because of the aforementioned differences between the real dielectric function and the one used in the model. Second, we examine changes in the peak intensity (Fig.~\ref{fig3}g). The peak loss probability increases with the NP size for both dipole and bulk plasmons in both the experiment and the simulations, with the slope being steeper for the bulk plasmon. Third, we study the quality factors (Q factors) of the resonances, defined as
\begin{align}
    Q = \frac{E_\mathrm{res}}{2\gamma}.
\end{align}
The resulting Q factors, shown in Fig~\ref{fig3}h, are rather low and reach values from 1.0 to 1.5. Compared to traditional plasmonic materials~\cite{West2010}, the Q factors of LSP in metallic VO$_2$ NPs are much lower. This is attributed to the dielectric properties of metallic VO$_2$, which is not a good metal in the sense of conductivity. However, it is a promising material for advanced electronics and optics due to its unique phase transition.

\subsection{Partially metallic VO$_2$ nanoparticles}

\begin{figure}[t!]
\includegraphics[width=1\textwidth]{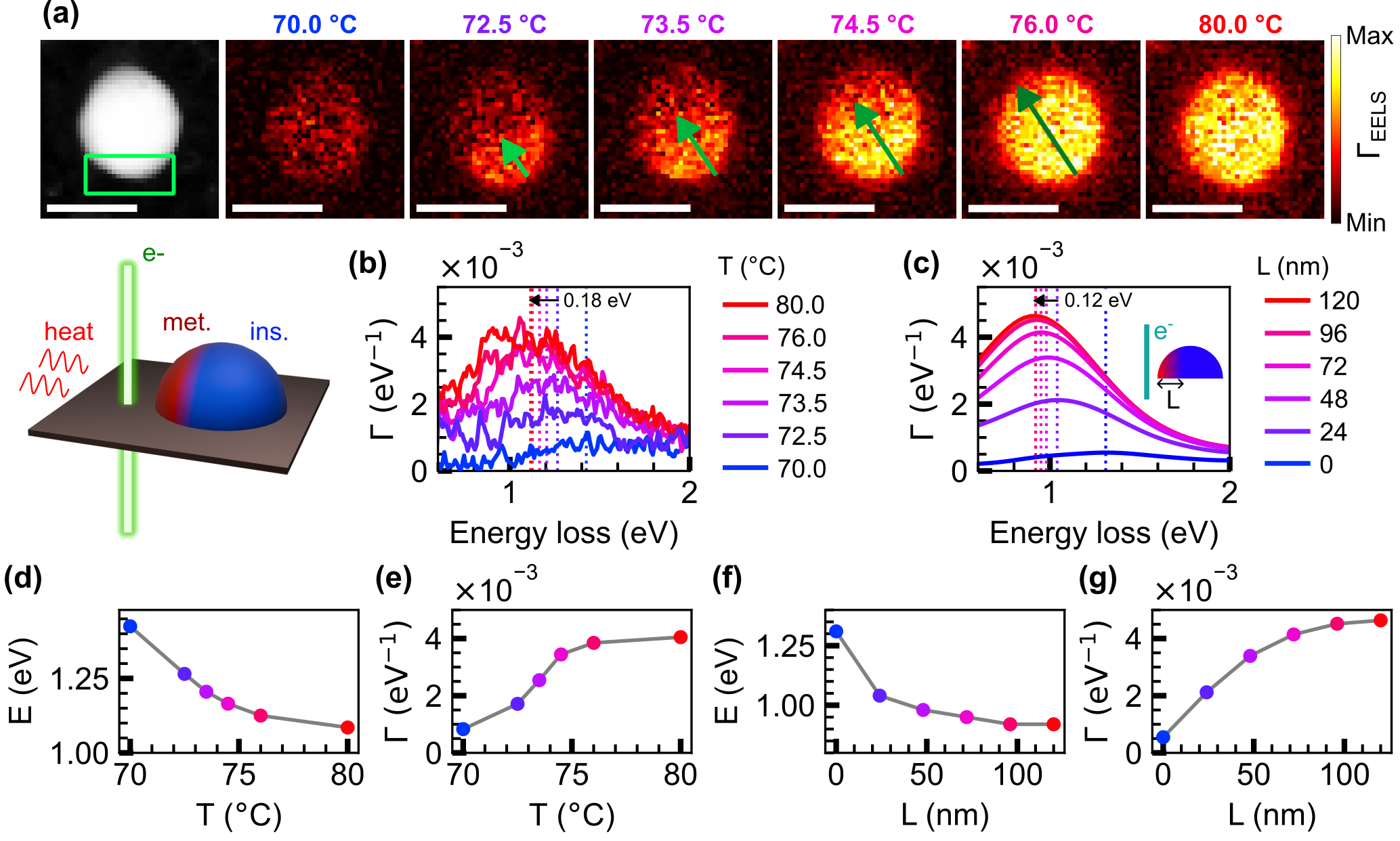}
\caption{Tuning of the dipole plasmon mode via partial insulator-metal transition. (a) HAADF STEM image of a partially switched NP and EEL maps (1.0--\SI{1.5}{\electronvolt}) for the NP measured at different temperatures (top) and the schematic depiction of the thermally tunable NP. The green arrows indicate the approximate size of the metallic part (brighter area).  (b) Measured EEL spectra for four transition states of the NP. (c) Simulated EEL spectra for \SI{120}{\nano\metre} hemisphere with a beam \SI{10}{\nano\metre} outside. We change the position of the boundary between the insulating and the metallic phase, increasing the size $L$ of the metallic part. (d,e) Experimental peak energy (d) and peak maxima (e) dependence on temperature. (f,g) Simulated peak energy (f) and peak maxima (g) as a function of the size $L$ of the metallic part.
\label{fig4}}
\end{figure}

So far, we have investigated the plasmonic properties of single-phase VO$_2$ nanoparticles. Nevertheless, plasmon modes have also been observed in partially switched nanoparticles, which opens the way to tune the LSP dynamically by the actual transition state. When the nanoparticles were heated and the EELS measurements were performed for different temperatures, we observed that some nanoparticles exhibited a gradual phase change. The behavior of such VO$_2$ NP with a diameter of \SI{120}{\nano\metre} is depicted in detail in Fig.~\ref{fig4}. Figure~\ref{fig4}a shows the HAADF STEM micrograph followed by EEL maps in an energy interval from 1.0 to \SI{1.5}{\electronvolt} at different temperatures covering the entire gradual phase change. An increase in the EEL intensity is observed from the lower right corner as a result of the excitation of plasmons in the growing metallic part, which is marked by green arrows in the partially switched states. To study the behavior of the plasmonic response, we examined the EEL spectra (Fig.~\ref{fig4}b), taken from the region highlighted in the HAADF image by the green rectangle, corresponding to the region where the phase transition starts. Looking at these spectra, we can see that for a low temperature (blue, \SI{70}{\celsius}), there is no evidence of a plasmon peak. When heating (purple, \SI{72.5}{\celsius}), the peak begins to emerge near \SI{1.3}{\electronvolt}, and after continuation of the heating, the peak increases in intensity and shifts towards lower energies. In the fully metallic state (red, \SI{80}{\celsius}), the peak has the highest intensity at an energy of \SI{1.1}{\electronvolt}. Consequently, the discussed VO$_2$ NP represents a system that can be switched on and off and tuned to an extent of \SI{0.18}{\electronvolt}.

To verify this behavior by theory, we replicate the system in simulations by splitting the hemisphere into two parts, one metallic and one insulating, and moving the boundary between them. By doing so, we effectively increase the volume of the metallic part by increasing its length $L$, resulting in a redshift of the dipole mode, and reduce the insulating part, which acts as a dielectric environment. The calculated spectra are summarized in Fig.~\ref{fig4}c. We observed a behavior similar to that in the experiment. There is no evidence of a plasmon peak in a fully insulating NP (blue, $L=0\,\mathrm{nm}$). With an increasing metallic part, the peak begins to emerge and further increases in intensity and shifts towards lower energies. In the fully metallic state (red, $L=120\,\mathrm{nm}$), the peak has the highest intensity and the lowest energy. The tunability of this model system is around \SI{0.12}{\electronvolt}. For better visibility of this behavior, we plot the peak energies and maxima values of the experimental peaks as a function of temperature in Fig.~\ref{fig4}d,e, respectively. Similarly, for the simulated peaks, we plot their energies and maxima values as a function of the length of the metallic part in Fig.~\ref{fig4}f,g.

As a result, the dynamics of the phase transition in the discussed VO$_2$ NP allowed us to establish a correlation between the local temperature and the relative switched volume. This correlation was achieved by employing the dipole plasmon mode in the metallic region as the linking variable, with the peak position serving as an indicator. This phenomenon can be generalized to VO$_2$ nanostructures with a higher aspect ratio that would introduce a wider tunability of the system. This will enable to locally detect the temperature and study the dynamics of the phase transition by optical methods.
In addition, it could be used to design an active optical element with a tunable operational wavelength, for example, near the telecommunication wavelength of \SI{1550}{\nano\metre} (\SI{0.8}{\electronvolt} in energy).

% =========================================================
% Conclusion
% =========================================================

\section{Conclusion}

In conclusion, we have studied an ensemble of VO$_2$ nanoparticles of hemispherical shape via STEM-EELS and explained the features in the loss spectra in the low-temperature insulating and high-temperature metallic phases, arising from the excitation of plasmons. A spatially-resolved study was performed on a single NP to study the origin of individual peaks spectrally overlapped due to high damping in the VO$_2$ material. With the help of numerical simulations, we confirm the nature of losses as LSP contributions, excited through the near-field when the electron beam passes outside the NP, and bulk contributions, when the electron travels through the VO$_2$. Furthermore, the size-tuning capabilities of the resonances were investigated. By processing the data, we extracted the energies, maximum intensities, and quality factors of the LSP and bulk resonances for the ensemble of NPs. The expected redshift of the LSP contribution was observed to occur in conjunction with the enlargement of the NP. Furthermore, a nearly constant energy of the bulk plasmon was observed, as it generally does not depend on the NP size. The energy tunability of the LSP is less pronounced in experiments, reading \SI{0.1}{\electronvolt} for the size of NPs from 50 to \SI{220}{\nano\metre}. However, it is substantial (\SI{0.32}{\electronvolt}) for the size range calculated in simulations (20 to \SI{500}{\nano\metre}). The quality factor of the VO$_2$ resonances was found to be significantly lower than that of traditional plasmonic materials, with values ranging from 1 to 1.5.

In addition, we have observed plasmon modes in partially switched NPs, with part of the sample remaining in the low-temperature (insulating) phase and part in the high-temperature (metallic) phase. With gradual heating, we have found that the metallic part grows and, subsequently, the plasmon mode redshifts. The discussed \SI{120}{\nano\metre}  VO$_2$ NP represents a system that can be switched on and off and tuned to the extent of \SI{0.18}{\electronvolt}. Moreover, it can be used to establish a correlation between the local temperature and the relative switched volume. Hence, we can conclude that this phenomenon is capable of being generalized to nanostructures that possess a higher aspect ratio, thereby resulting in a wider degree of tunability of the system. We envision that such thermal tunability may be crucial in the development of VO$_2$ nanodevices.

% =========================================================
% Methods
% =========================================================

\section{Methods}\label{Methods}

\subsection{Sample preparation}\label{MethodsSample}
VO$_2$ nanoparticles were manufactured on a heating chip (Protochips) with a \SI{30}{\nano\metre} thick silicon nitride membrane in a two-step process introduced in \cite{Kepic2025}. First, a \SI{30}{\nano\metre} thick amorphous film was deposited by a TSST pulsed laser deposition (PLD) system using the following parameters: \SI{248}{\nano\metre} KrF laser, \SI{2}{\joule\per\square\centi\metre}, \SI{10}{\hertz}, 50000 pulses, vanadium target (99.9\% purity, Mateck GmbH), \SI{50}{\milli\metre} substrate-target distance, room temperature, and 5\,mTorr oxygen pressure. Second, the sample was ex situ annealed for 30 minutes in a vacuum furnace (Clasic CZ Ltd.) at \SI{700}{\celsius} under 15\,sccm oxygen flow to dewet the amorphous film into VO$_2$ nanoparticles.

\subsection{TEM measurements}\label{MethodsTEM}

STEM EELS measurements (Figures~\ref{fig1} to \ref{fig3}) were performed on the Nion aberration-corrected high energy resolution monochromated EELS-STEM (HERMES$^{\mathrm{TM}}$) equipped with a prototype Nion spectrometer possessing a Hamamatsu ORCA high-speed CMOS detector and an in-situ Fusion Select system (Protochips) for heating experiments at Oak Ridge National Laboratory. The primary beam energy was set at \SI{60}{\kilo\electronvolt} to suppress relativistic losses such as Čerenkov radiation \cite{Horak2015} in low-loss EELS. The electron beam current was around \SI{300}{\pico\ampere}. The convergence semi-angle in STEM was set at \SI{15}{\milli\radian}. The HAADF-STEM signal was collected using scattered electrons at an angle of \SI{90}{\milli\radian} to \SI{200}{\milli\radian} and the EELS collection angle was \SI{13}{\milli\radian}.

Part of the TEM analysis (Figure~\ref{fig4}) was performed on a TEM FEI Titan equipped with a monochromator, a GIF Quantum spectrometer for EELS, and an in-situ Fusion Select system (Protochips) for heating experiments at CEITEC Nano. The primary beam energy was set at \SI{120}{\kilo\electronvolt} to achieve an optimal signal-to-background ratio \cite{Horak2020}. The electron beam current was around \SI{100}{\pico\ampere}. The convergence semi-angle in STEM was set at \SI{8.14}{\milli\radian}. The ADF-STEM signal was collected using a Gatan ADF detector that captures scattered electrons at an angle of \SI{16.7}{\milli\radian} to \SI{38.3}{\milli\radian}, and the EELS collection angle was \SI{8.3}{\milli\radian}.

EEL spectra were integrated over the marked regions of interest to reduce noise. They were further divided by the integral intensity of the zero-loss peak to transform the measured counts into a quantity proportional to the loss probability. Finally, the EEL spectrum of a pure silicon nitride membrane was subtracted to remove the background. The processing of EEL spectra is shown in detail in the Supplementary Information (Fig.~\ref{SI_fig1}).

\subsection{Theoretical calculations}\label{MethodsSimulations}

Electron energy-loss spectra of the hemispheres were numerically calculated in COMSOL Multiphysics, using classical dielectric formalism~\cite{Abajo2010} and the approach introduced in Ref.~\cite{Konecna2018}. The dielectric function of VO$_2$ was taken from Ref.~\cite{Kepic2025}. We utilize a non-recoil approximation for the electron beam and implement it by a current flowing through a line segment parallel with the $z$ direction and passing through the position $\mathbf{R}_\mathrm{b}$ in the $xy$-plane. The loss probability function reads
\begin{align}
\Gamma_\mathrm{EELS}(\omega)= \frac{e^3}{\pi\hbar^2\omega}\int\mathrm{d}z\,\Re\Big\{ E_z^{\mathrm{ind}}(\mathbf{R}_\mathrm{b},z,\omega)\mathrm{e}^{\mathrm{-i}\omega z / v}\Big\} ,
\end{align}
where $E_z^{\mathrm{ind}}$ is the component of the induced electric field parallel to the electron-optical axis and the integration is performed on the trajectory of the electron beam, $v$ the electron velocity, and $\hbar\omega$ the energy loss. We assumed that the particle was positioned in a free space. The coupling effect between the nanoparticles was neglected because the distances between the particles are tens or hundreds of nanometers.  

The retarded bulk loss function for an electron passing through a medium reads~\cite{Abajo2010}
\begin{align}
\Gamma_\mathrm{bulk}(\omega)
= \frac{e^2L}{4\pi^\mathrm{2}\hbar\varepsilon_0 v^2}\Im\left\{\left(\frac{v^2}{c^2}-\frac{1}{\varepsilon_\mathrm{r}(\omega)}\right)\mathrm{ln}\left(\frac{q_\mathrm{c}^2-(\omega/c)^2\varepsilon_\mathrm{r}(\omega)}{(\omega/v)^2-(\omega/c)^2\varepsilon_\mathrm{r}(\omega)}\right)\right\},
\label{eq_Bulk_loss}
\end{align}
where $L$ is the distance that the electron passes through the material volume, $c$ the speed of light, $\varepsilon_\mathrm{r}$ the dielectric function of VO$_2$ taken from Ref.~\cite{Kepic2025}, and the wave vector $q_\mathrm{c}$ corresponds to the momentum cutoff $\hbar q_\mathrm{c}\approx\sqrt{\left(m_\mathrm{e}v\phi_\mathrm{out}\right)^2+\left(\hbar\omega/v\right)^2}$, below which momentum transfers are collected and is determined by the half-aperture collection angle of the microscope $\phi_\mathrm{out}$. The values were set at $v = 0.446\,c$ (primary beam energy of \SI{60}{kV}) and $\phi_\mathrm{out} =$ \SI{13}{\milli\radian}. 

% =========================================================
% Acknowledgements, SI available, References
% =========================================================

\begin{acknowledgement}

This work has been supported by the Czech Science Foundation (grants No. 22-04859S and 23-05119M), the OP JAK Excellent Research program (project QM4ST, No. CZ.02.01.01/ 00/22\_008/0004572),  MEYS CR (project Czech-NanoLab, No. LM2023051) and Brno University of Technology (grants No. FSI-S-23-8336 and CEITEC VUT-J-25-8860). Ultrahigh energy resolution monochromated EELS was performed at the Center for Nanophase Materials Sciences (CNMS), which is a US Department of Energy, Office of Science, User Facility, using instrumentation within ORNL's Materials Characterization Core provided by UT-Battelle, LLC, under Contract No. DE-AC05- 00OR22725 with the DOE and sponsored by the Laboratory Directed Research and Development Program of Oak Ridge National Laboratory, managed by UT-Battelle, LLC, for the U.S. Department of Energy. J.K. acknowledges the Brno Ph.D. Talent Scholarship -- Funded by the Brno City Municipality.

\end{acknowledgement}

\bibliography{bibliography}

\newpage
\section{Supplementary Information}
\renewcommand{\thefigure}{S\arabic{figure}}
\setcounter{figure}{0}

\begin{figure}[h]
\includegraphics[width=1\textwidth]{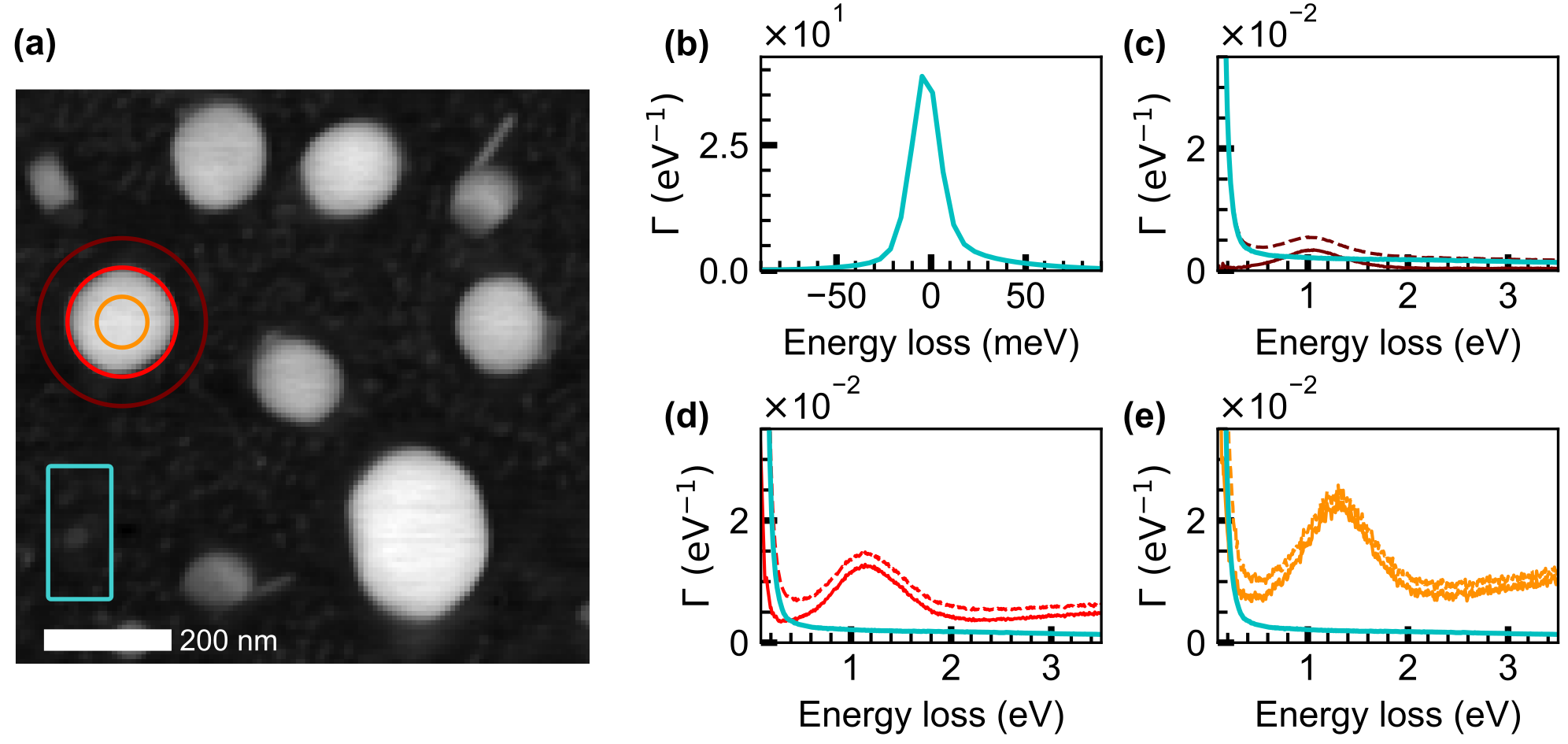}
\caption{Processing of EEL spectra. (a) HAADF STEM micrograph with marked integration areas for center (orange), edge (red), and out (maroon) signal as well as for the background, i.e., the signal from a pure silicon nitride membrane (cyan). (b) Typical zero-loss peak proving the high energy resolution of our experiments. (c) Full EEL spectrum for the out position (dashed maroon) overlaid by the background (cyan) and background-subtracted EEL spectrum (solid maroon). (d) Full EEL spectrum for the edge position (dashed red) overlaid by the background (cyan) and background-subtracted EEL spectrum (solid red). (e) Full EEL spectrum for the center position (dashed orange) overlaid by the background (cyan) and background-subtracted EEL spectrum (solid orange). Within the manuscript, background-subtracted EEL spectra are presented.
\label{SI_fig1}}
\end{figure}

\newpage
\begin{figure}[h]
\includegraphics[width=1\textwidth]{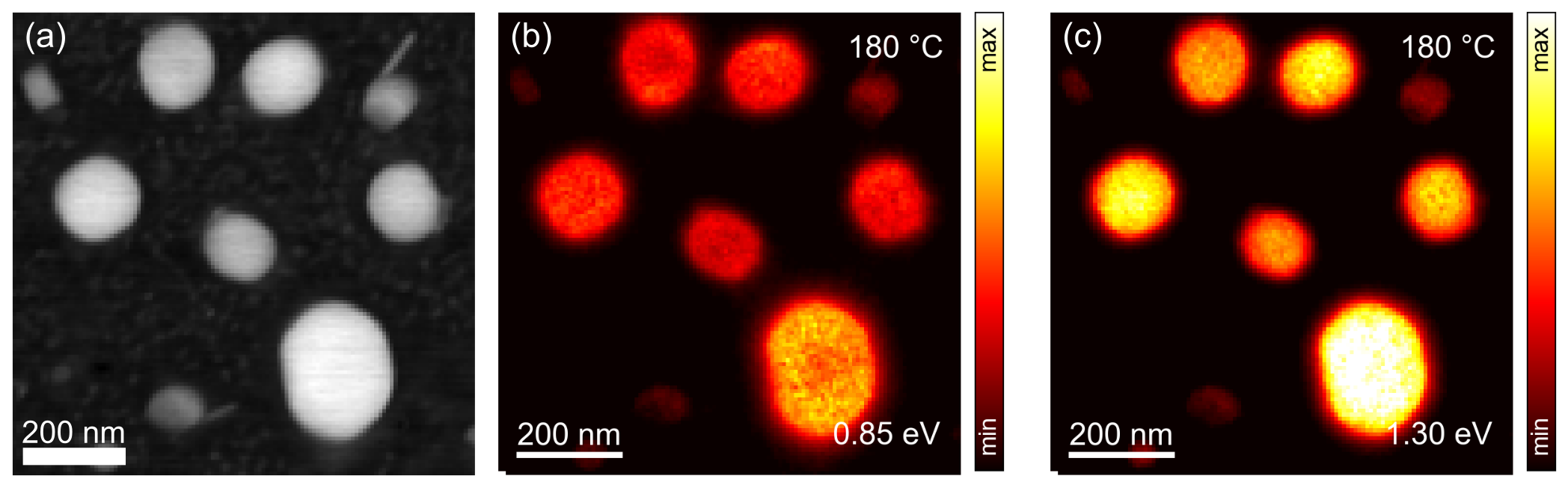}
\caption{Energy filtered loss probability maps of VO$_2$ NPs. (a) HAADF STEM image of one region of the ensemble of NPs. (b) Loss probability map at the energy of \SI{0.85}{\electronvolt}. Even though the thickness is smaller at the edges compared to the center, the signal is uniform over some NPs or highest at the edges for others, indicating the dominant contribution of localized surface plasmons (the dipole mode). (c) Loss probability map at the energy \SI{1.3}{\electronvolt}, where we observe hotspots at the center, where the nanoparticles are thicker, confirming the dominant excitation of bulk plasmon.
\label{SI_fig2}}
\end{figure}

\newpage
\begin{figure}[h]
\includegraphics[width=1\textwidth]{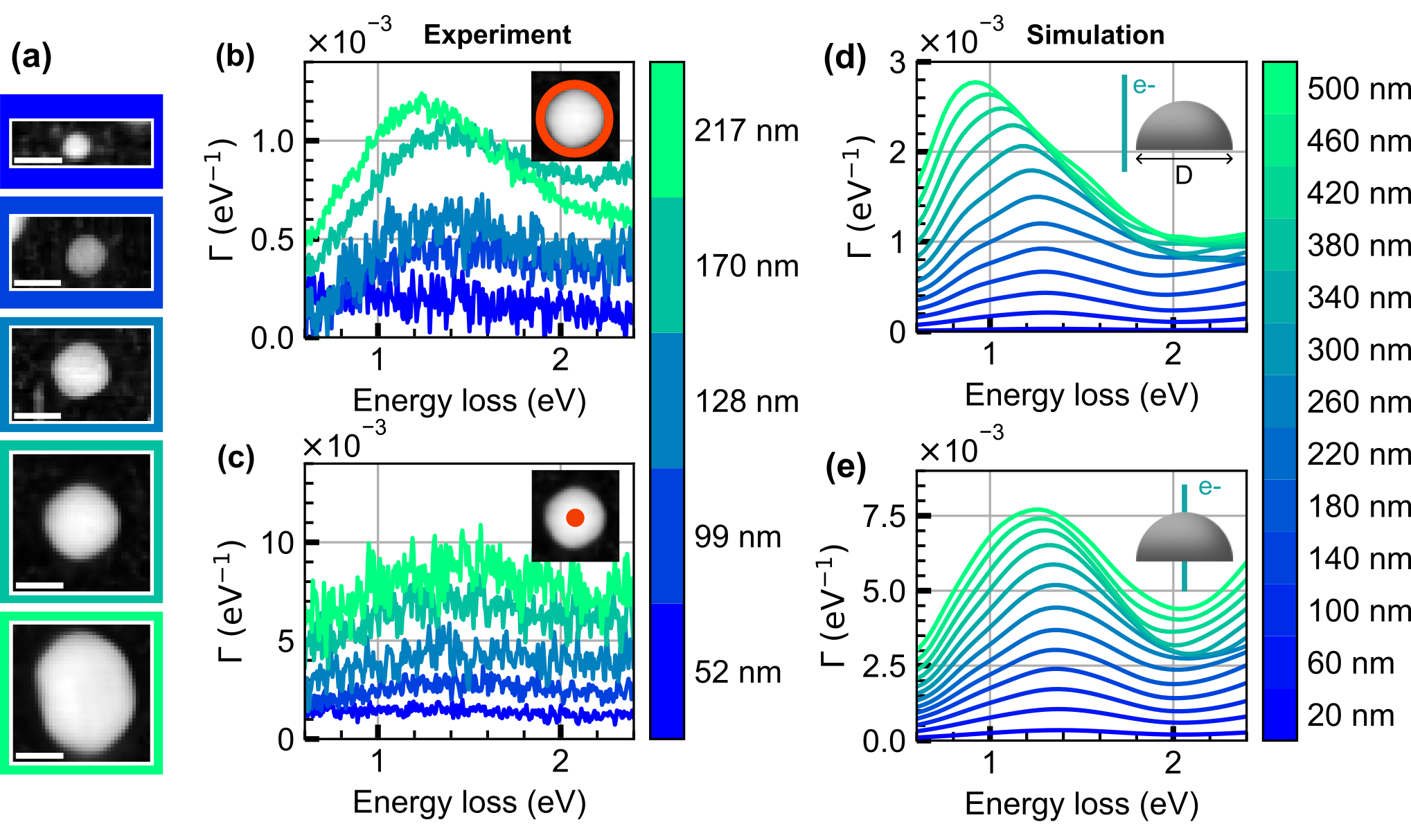}
\caption{Size dependence of EEL spectra for insulating VO$_2$ NPs. (a) HAADF STEM microcrographs of five representatives. The scalebars correspond to \SI{100}{\nano\metre}. (b) Measured EEL spectra taken from the region outside the NPs. (c) Measured EEL spectra taken from the center of the NPs. (d) Simulated EEL spectra for beam position \SI{10}{\nano\metre} from the outer edge of the NPs. (e) Simulated EEL spectra for the electron beam positioned in the center. The sizes of the NPs are at the individual colorbar legends. In the case of edge excitation (b,d), we observe for the smaller NPs an increase in signal with increasing their size. Furthermore, an energy shift appears for the largest NPs, indicating additional excitation of localized resonances in the dielectric. For the central position (c,e), an increase in signal is observed with the increasing NP size.
\label{SI_fig3}}
\end{figure}

\end{document}